\documentclass[12pt]{article}
 \usepackage{amsmath}
 \usepackage{amssymb}
 \usepackage{epsfig}
\begin{document}
 \begin{center}
 {\Large \bf Group-theoretic Approach for
 Symbolic Tensor Manipulation: II. Dummy Indices}
 \end{center}

 \centerline{L. R. U. Manssur\footnote{Email: leon@cbpf.br} and
 R. Portugal\footnote{Email: portugal@lncc.br}}

 \begin{center}
 Laborat\'orio Nacional de Computa\c{c}\~ao Cient\'{\i}fica,\\
 Av. Get\'ulio Vargas, 333,\\
 Petr\'opolis, RJ, Brazil. Cep 25651-070.
 \end{center}

 \begin{abstract}
 Computational Group Theory is applied to indexed objects
 (tensors, spinors, and so on) with dummy indices. There are
 two groups to consider: one describes the intrinsic
 symmetries of the object and the other describes the
 interchange of names of dummy indices. The problem of
 finding canonical forms for
 indexed objects with dummy indices reduces to finding double
 coset canonical representatives.
 Well known computational group algorithms are applied to
 index manipulation, which allow to address the simplification
 of expressions with hundreds of indices going
 further to what is needed in practical applications.

 \

 {\bf Key words.} Symbolic tensor manipulation, Computational
 Group Theory, Algorithms, Canonical coset representative,
 Symmetric group\footnote{2000
 Mathematics Subject Classification. 70G45,
 20B40, 53A45, 20B35, 53A35, 20B30, 53A15}

 \end{abstract}

 \section{Introduction}

 Ref. \cite{Benar} describes how Computational
 Group Theory provides
 tools for manipulating tensors with free indices. The tensors
 obey what we call {\it permutation symmetries}, which are a
 set of tensor equations of the form
 \begin{equation}
 T\,^{{i_{1}}}\,^{\cdots}\,^{{i_{n}}}=\epsilon_\sigma \,T\,^{\sigma (
 {i_{1}}\,\cdots\,{i_{n}})},
 \label{Ts}\end{equation}
 where $\sigma ({i_{1}}\,\cdots\,{i_{n}})$ is a permutation of
 ${i_{1}}\,\cdots\,{i_{n}}$ and
 $\epsilon_{\sigma}$ is either 1 or $-1$.  This kind of symmetry
 can be described by finite group theory and the index manipulation can be
 performed using the algorithms of Computational Group Theory
 \cite{Sims,Leon,Butler1,Butler2,Hoffmann,Butler_book}. The detailed
 description of symmetry as a group is given in ref. \cite{Benar}.

 In this work we address the problem of applying Computational
 Group Theory for manipulating dummy indices.
 It is more complex then the
 free index problem, since one has to deal with two groups:
 the group that describes the symmetries of the indexed object and
 the group that describes the symmetry of interchange of dummy
 indices. These groups act on a standard index configuration,
 generating sets of equivalent configurations. These sets are
 double cosets, which have already been studied in
 Computational Group Theory \cite{Butler2,Laue,Conway}.
 The most important concept for simplifying tensor
 expressions is the determination of canonical forms,
 which correspond to canonical representatives of
 single cosets for free indices and
 double cosets for dummy indices.
 The algorithms of the present work and the algorithms
 of refs. \cite{Benar}, \cite{Portugal1}, and \cite{Portugal2}
 allow the manipulation of expressions built out of indexed
 objects obeying permutation symmetries,
 such as tensors, spinors, objects with gauge indices,
 and so on, with commutative or anticommutative properties.
 On the other hand, these algorithms do not solve yet
 the problem when there are algebraic constraints, such as
 the cyclic symmetry of the Riemann tensor.

 Manipulation of dummy indices can also be found
 in general algebraic expressions with sums and
 multiple integrals. For instance, the calculation of
 Feynman diagrams in Quantum Field Theory generates a large
 number of multiple integrals of the propagator which
 can in principle be reduced using the algorithms of this
 work by canonicalizing the integration variables.


 The structure of this paper is as follows. In section 2
 we describe the representation theory
 for dummy indices. In section 3 we describe the algorithm to
 canonicalize indexed objects with dummy indices, and
 in section 4 we discuss the algorithm complexity.
 In section 5 we discuss the simplification of
 general expressions in order to have a bird's-eye view
 of the problem, and present an example of canonicalizing
 a Riemann monomial of degree 3.

 We assume that the reader is familiar with the concepts and
 notations described in ref. \cite{Benar}.

 \section{Representation theory for dummy indices}

 Suppose that $T$ is a fully contracted rank-$2n$ tensor
 with symmetry $S$.
 We define the standard configuration as
 \begin{equation}
 T^{d_1}\,_{d_1}\,^{d_2}\,_{d_2}\,\cdots\,^{d_n}\,_{d_n}.
 \label{abs}
 \end{equation}
 This configuration is associated with permutation $+id$, which
 is the least element of the symmetric group $S_{2n}$.
 Our first task is to determine the configurations that are equivalent to
 (\ref{abs}). We know that the dummy index names can be interchanged.
 For example, the configuration
 \begin{equation}
 T^{d_2}\,_{d_2}\,^{d_1}\,_{d_1}\,\cdots\,^{d_n}\,_{d_n}
 \label{e2}
 \end{equation}
 is equivalent to (\ref{abs}) and is obtained by the action of the element
 $+(1,3)(2,4)$ on (\ref{abs}). This element is not in $S$ in general.
 If the metric is symmetric, the configuration
 \begin{equation}
 T_{d_1}\,^{d_1}\,^{d_2}\,_{d_2}\,\cdots\,^{d_n}\,_{d_n}
 \label{e3}
 \end{equation}
 is also equivalent to  (\ref{abs}) and is obtained by the action of the element
 $+(1,2)$. What is the group that describes
 these kinds of symmetries?
 Let $D$ be a subgroup of $H\otimes S_{2n}$ generated by
 \begin{eqnarray}
 K_D = \{\,+(1,2),\,+(3,4),\,\cdots,\,+(2n-1,2n),\hspace{5.0cm} \nonumber\\
 +(1,3)(2,4),\,+(3,5)(4,6),\,\cdots,\,+(2n-3,2n-1)(2n-2,2n)\},
 \label{e4}
 \end{eqnarray}
 with the base $\boldsymbol b_D=[1,3,\cdots,2n-1]$. $K_D$ is a strong
 generating set with respect to $\boldsymbol b_D$. The action of $D$ on
 configuration
 (\ref{abs}) yields all configurations that can be obtained
 from (\ref{abs}) by interchanging dummy index names or
 by using the symmetry of the metric.

 Besides the action of $D$, we consider the action of $S$.
 If we take
 configuration (\ref{abs}) as the starting point, similar as we have
 done for the free index case, we have to apply $D$ first, followed
 by $S$ in order to obtain all configurations equivalent to (\ref{abs}).
 This order is crucial.  If one
 applied an element of $S$ first, the positions of the dummy
 indices  would change and the application of $D$ on this new configuration
 would make no sense. It would not be an interchange of dummy index names nor
 an interchange of contravariant index to a covariant one inside a
 pair. Let us see an example. Suppose that $S=\{+id,\,+(2,3)\}$,
 and let us apply $+(2,3)$ on configuration (\ref{abs})
 followed by $+(3,5)(4,6)\in D$. We obtain
 \begin{equation}
 T^{d_1}\,^{d_2}\,^{d_3}\,_{d_3}\,_{d_1}\,_{d_2}\,\cdots\,^{d_n}\,_{d_n},
 \label{e6}
 \end{equation}
 which is not equivalent to (\ref{abs}) at all. The reverse order
 is perfectly fine, let us apply  $+(3,5)(4,6)$ first, followed
 by $+(2,3)$:
 \begin{equation}
 T^{d_1}\,^{d_3}\,_{d_1}\,_{d_3}\,^{d_2}\,_{d_2}\,\cdots\,^{d_n}\,_{d_n}.
 \label{e7}
 \end{equation}
 The configuration above is equivalent to (\ref{abs}).

 The set ($\mathcal C$) of all configurations equivalent to (\ref{abs})
 is given by the action of $S\times D$ on (\ref{abs}), i.e.
 \begin{equation}
 \mathcal C = \{(s\, d)(
 T^{d_1}\,_{d_1}\,^{d_2}\,_{d_2}\,\cdots\,^{d_n}\,_{d_n}),
 \,s\in S,\,d\in D\}.
 \label{C}
 \end{equation}
 The set  $S\times D$ is the double coset
 of $S$ and $D$ in  $H\otimes S_{2n}$ that contains the
 identity $+id$. The cardinality of this set is
 $|S||D|/|S\cap D|$.

 Consider a fully contracted
 configuration ($T_1$) that is not equivalent to (\ref{abs}),
 one can take (\ref{e6}) as an example.
 Suppose that
 $T_1$ is obtained by acting $g$
 on (\ref{abs}).
 Then $g \not\in S\times D$.
 The set of all configurations
 equivalent to $T_1$ is given by the action of the double
 coset $S\times g\times D$ on (\ref{abs}).
 The cardinality of this set is
 $|S||D|/|S^g\cap D|$, where $S^g$ is the conjugate
 set $g^{-1} \times S \times g$ \cite{Hoffmann}.

 \section{Algorithm to canonicalize tensors with dummy indices}

 Now we address the following problem.
 Suppose one gives a fully contracted rank-$2n$ tensor which
 has some symmetry described by a set of tensor equations of the
 form (\ref{Ts}).
 Find the canonical index configuration using the tensor symmetries,
 renaming of dummy indices, and the metric symmetries.

 In representation theory, this problem can be solved
 if one knows the solution of the following equivalent problem.
 Given a generating set $K_S$ for the group $S$,
 and an element $g=(\epsilon_\pi,\pi)\in H\otimes S_{2n}$,
 find the canonical representative
 of the double coset $S\times g\times D$,
 where $D$ is the group generated by (\ref{e4}) with
 respect to the base $\boldsymbol b_D$.

 Butler \cite{Butler2} describes an algorithm for determining the
 double coset canonical representative for permutations
 groups. The input of his algorithm is:

 \

 \noindent
 (a) a permutation group $G$ acting on a set $P$ with a base
 $\boldsymbol b=[b_1,\cdots,b_k]$;\\
 (b) subgroups $A$ and $B$ of $G$ given by a base and
 strong generating set; and\\
 (c) an element $g$ of $G$.

 \

 \noindent
 The algorithm determines the image of the base $\boldsymbol b$
 under the first element of the double coset $A\times g\times B$.

 We modify Butler's algorithm in order to work
 within the direct product $H\otimes S_{2n}$.  Fortunately,
 in the tensor problem, the subgroup $D$ is fixed and we know
 beforehand a base and a strong generating set for it.
 Butler's algorithm keeps changing the base for the subgroup $B$
 (see item (b) above) during the determination of the image of the
 canonical representative. This base change is very simple for
 group $D$.

 Suppose that the settings in terms of tensor notation have already
 been converted into group notation. So, the input
 of the algorithm is:

 \

 \noindent
 (a) $n$;\\
 (b) a base $\boldsymbol b_S=[b_1,\cdots,b_k]$ and strong generating set $K_S$
  for $S$; and\\
 (c) an element $g=(\epsilon_\pi,\pi)\in H\otimes S_{2n}$.

 \

 \noindent
 The output is either
 the canonical representative $\bar{g}=(\epsilon_{\bar{\pi}},\bar{\pi})$
 of the double coset $S\times g \times D$ or 0. The output
 0 occurs if and only if both $(\epsilon_{\bar{\pi}},\bar{\pi})$ and
 $(-\epsilon_{\bar{\pi}},\bar{\pi})$ are in $S\times g \times D$.
 To have the solution in terms of tensor notation when the output
 is not 0, one simply acts $\bar{g}$ on (\ref{abs}).

 The algorithm basically consists of $2n-1$ loops. We describe the first two
 loops, which are enough to understand the whole process.
 The algorithm is formally described ahead. The base
 $\boldsymbol b_S$ must be extended in order to be
 a base for $S_{2n}$. Let
 $\boldsymbol b_S=[b_1,\cdots,b_k,b_{k+1},\cdots,b_{2n-1}]$
 be the extended base and $[p_1,\cdots,p_{2n-1}]$ the image
 of the canonical representative. The goal is to
 find $p_1,\cdots,p_{2n-1}$ each at a time. First loop determines $p_1$.

 The orbit $b_1^{S\times g \times D}$
 gives all possible values of the first point of the image of the
 elements of the double coset
 $S\times g\times D$. Call this set IMAGES$_1$.
 $p_1$ is the least point of IMAGES$_1$ with respect
 to $\boldsymbol b_S$. So,
 if $\Delta_{b_1}=b_1^S$ then
 \begin{equation}
 {\rm IMAGES}_1=\bigcup\limits_{i\in (\Delta_{b_1})^{g}} i^D.
 \label{p1}
 \end{equation}
 The least point is calculated with respect to base
 $\boldsymbol b_S$. The order of points is $b_1< \cdots <b_{2n}$.
 Before finding $p_2$, we have to determine the pairs
 of elements $(s_1,d_1)$ of $(S,D)$ that satisfy
 \begin{equation}
 b_1^{s_1\, g \, d_1}=p_1,
 \label{b1}
 \end{equation}
 since, from now on, $p_1$ must remain as the first point.
 Suppose that $(s_1,d_1)$ is a pair that satisfies (\ref{b1}),
 then
 \begin{equation}
 b_1^{S_{b_1}\times s_1\, g \, d_1\times D_{p_1}}=p_1.
 \label{b1a}
 \end{equation}
 So we have to determine a small set of pairs $(s_1,d_1)$ and
 amplify it by using the stabilizers $S_{b_1}$
 and $D_{p_1}$ in order to obtain all pairs $(s_1,d_1)$ that
 satisfy (\ref{b1}).
 Notice that to determine $D_{p_1}$,
 we have to perform a base change so that
 $p_1$ becomes the first point of $D$ .
 This is easily performed.
 The pairs $(s_1,d_1)$
 are stored in table TAB defined in the following way:
 \begin{equation}
 {\rm TAB}([i])=(s_1,d_1),\,[i]\in {\rm ALPHA}_1,
 \label{TAB}
 \end{equation}
 where ALPHA$_1$ is defined by
 \begin{equation}
 {\rm ALPHA}_1=\{[i],\, i \in b_1^S\,{\rm and}\,i^g\in p_1^D\}.
 \label{A}
 \end{equation}
 The size of list $[i]$ is given by the index 1 of ALPHA$_1$.
 One can find ALPHA$_1$ using
 \begin{equation}
 {\rm ALPHA}_1=\Delta_{b_1} \cap (p_1^D)^{g^{-1}}.
 \label{ALPHA1}
 \end{equation}
 We omit the dependence of $(s_1,d_1)$ on the entries of ALPHA$_1$,
 since the explicit
 notation $(s_1([i]),d_1([i]))$ is cumbersome.
 It is important to keep in mind that
 for each entry of ALPHA$_1$ there is a correspondent
 pair $(s_1,d_1)$. Note that the variables with index 1
 are calculated in the first loop of the algorithm.

 The pair $(s_1,d_1)$ corresponding to $[i]$ is given by
 \begin{eqnarray}
 s_1=trace(i,\nu_S),\hspace{0.5cm}
 \nonumber \\
 d_1=trace(i^g,\nu_D)^{-1},
 \label{d1}
 \end{eqnarray}
 where $\nu_S$ and $\nu_D$ are the Schreier vectors relative to
 the orbits of $S$ and $D$ respectively.
 First loop finishes here.

 For each pair $(s_1,d_1)$ we
 calculate $b_2^{S_{b_1}\times s_1\, g \, d_1 \times D_{p_1}}$
 in order to obtain
 \begin{equation}
 {\rm IMAGES}_2 = \bigcup\limits_{[i]\in {\rm ALPHA}_1}
 \left ((b_2^{S_{b_1}})^{s_1\, g \, d_1} \right )^{D_{p_1}}.
 \label{IMAGES2}
 \end{equation}
 IMAGES$_2$ yields all images of $b_2$ in the double coset
 $S\times g\times D$ obeying the constraint (\ref{b1}). Then
 $p_2$ is the least point of IMAGES$_2$.

 Now we show how to find ALPHA$_2$ and the
 associated pairs $(s_2,d_2)$. At this point, a
 pair $(s_2,d_2)$ has the following property:
 \begin{equation}
 [b_1,b_2]^{{s_2}\, g \, {d_2}} = [p_1,p_2].
 \label{b1b2}
 \end{equation}
 For each pair $(s_2,d_2)$, define
 \begin{equation}
 {\rm NEXT}_2=(b_2^{S_{b_1}})^{s_2}\cap (p_2^{D_{p_1}})^
 {(g\,d_2)^{-1}}.
 \label{NEXT2}
 \end{equation}
 This is the set of images of $b_2$ in $S$ that
 yields $p_2$ after applying $g$ and $d_2$. This set gives the points
 that extend ALPHA$_1$. So
 \begin{equation}
 {\rm ALPHA}_2=\{[i,j],\, [i] \in {\rm ALPHA}_1,\,
 j\in {\rm NEXT}_2\}.
 \label{Aij}
 \end{equation}
 For each $[i,j]$ in ALPHA$_2$ we have to determine
 a pair $(s_2,d_2)$ that satisfies
 \begin{equation}
 [b_1,b_2]^{S_{b_1,b_2}\times {s_2} \, g \, {d_2} \times D_{p_1,p_2}} =
 [p_1,p_2].
 \label{b1b2_2}
 \end{equation}
 Let
 \begin{eqnarray}
 s_2=trace(j^{s_1^{-1}},\nu_S^{(2)})\times {s_1},\hspace{0.2cm}
 \nonumber \\
 d_2={d_1}\times trace(j^{g\,d_1}, \nu_D^{(2)})^{-1},
 \label{d2}
 \end{eqnarray}
 where $\nu_S^{(2)}$ and $\nu_D^{(2)}$ are the Schreier vectors relative to
 the orbits of the stabilizers $S^{(2)}$ and $D^{(2)}$ respectively.
 We define the new entries of TAB as
 \begin{equation}
 {\rm TAB}([i,j])=(s_2,d_2),\,[i,j]\in {\rm ALPHA}_2,
 \label{TAB2}
 \end{equation}
 and clear the old ones. Second loop finishes here.

 In the $i$th loop, IMAGES$_i$ is given by
 \begin{equation}
 {\rm IMAGES}_i=\bigcup\limits_{L\in ALPHA}\,
 \left((b_i^{S^{(i)}})^{s_i\,g\,d_i} \right )^{D^{(i)}},
 \label{IM}
 \end{equation}
 where $s_i$ and $d_i$ are obtained from TAB($L$). NEXT$_i$ is given by
 \begin{equation}
 {\rm NEXT}_i=(b_i^{S^{(i)}})^{s_i}\cap (p_i^{D^{(i)}})^{(g\,d_i)^{-1}},
 \label{NE}
 \end{equation}
 and the pairs $(s_i,d_i)$ obey
 \begin{equation}
 [b_1,\cdots,b_i]^{s_i\,g\,d_i}=[p_1,\cdots,p_i].
 \label{sidi}
 \end{equation}

 Now we present algorithm Canonical for dummy indices and the
 sub-routines $F_1$ and $F_2$. We use a pseudo-language that can be
 converted into programs of some computer algebra
 system.

 \

 \centerline{\bf Algorithm Canonical (dummy indices)}

 \begin{tabbing}
 \={\bf procedure} \ double\_coset\_can\_rep\,($n$,\,$\boldsymbol
 b_S$,\,$K_S$,\,$g$)\\
 \\
 {\bf input:} \= $n$ number of pairs of dummy indices;\\
 \> $\boldsymbol b_S=[b_1,\cdots,b_k]$ base of group $S$;\\
 \> $K_S$ strong generating set of $S$ with respect to base $\boldsymbol b_S$;
 and\\
 \> an element $g=(\epsilon_\pi,\pi)\in H\otimes S_{2n}$. \\
 \\
 {\bf output:} \= $\bar{g}=(\epsilon_{\bar{\pi}},\bar{\pi})$
 canonical representative of the double coset $S\times g \times D$ or\\
 \> 0 if $(\epsilon_{\bar{\pi}},\bar{\pi})$ and
 $(-\epsilon_{\bar{\pi}},\bar{\pi})$ are in  $S\times g \times D$.\\
 \hspace{0.5cm}\=\hspace{0.9cm}\=\hspace{0.9cm}\=\hspace{0.9cm}\=
 \hspace{0.9cm}\=\hspace{0.9cm}\= \\
 {\bf begin} \\
 \> $\boldsymbol b_S\,:=\,[b_1,\cdots,b_k,b_{k+1},\cdots,b_{2n-1}]$
 is the extension of $\boldsymbol b_S$ in order\\
 \> \hspace{1.2cm} to be a base for $S_{2n}$; \\
 \> if metric is symmetric then\\
 \> \> $K_D\,:=\, \{\,$\=$+(1,2),\,+(3,4),\,\cdots,\,+(2n-1,2n),$\\
 \> \> \> $+(1,3)(2,4),\,+(3,5)(4,6),\,\cdots,\,+(2n-3,2n-1)(2n-2,2n)\}$;\\
 \> else\_if metric is antisymmetric then \\
 \> \> $K_D\,:=\, \{\,-(1,2),\,-(3,4),\,\cdots,\,-(2n-1,2n),$\\
 \> \> \> $+(1,3)(2,4),\,+(3,5)(4,6),\,\cdots,\,+(2n-3,2n-1)(2n-2,2n)\}$;\\
 \> else \\
 \> \>
 $K_D\,:=\,\{+(1,3)(2,4),\,+(3,5)(4,6),\,\cdots,\,+(2n-3,2n-1)(2n-2,2n)\}$;\\
 \> end if; \\
 \> $\boldsymbol b_D$\,:=\,$[1,\,3,\,\cdots,\,2n-1]$; \\
 \> $(*$ Initialize table TAB and ALPHA $*)$ \\
 \> TAB([\,])\,:=\,$(+id,+id)$;\\
 \> ALPHA$\,:=\,\{[\,]\}$; \\
 \\
 \> for $i$ from 1 to $2\,n-1$ do \\
 \> \> $\Delta_S\,:=\,$all orbits of $S$ (and calculate $\nu_S$ - Schreier vector
 with respect to $\boldsymbol b_S$);\\
 \> \> $\Delta_b\,:=\,b_i^{<K_S>}$; \\
 \> \> $\Delta_D\,:=\,$ all orbits of $K_D$; \\
 \> \> $(*$ IMAGES is given by eq. (\ref{IM}) $*)$\\
 \> \> IMAGES\,:=\,map $F_1$ on each entry of ALPHA passing TAB, \\
 \> \> \> \> \hspace{0.5cm}$\Delta_D$, $\Delta_b$ and $g$ as extra arguments; \\
 \> \> $p_i\,:=\,$least point of IMAGES with respect to base $\boldsymbol b_S$;
 \\
 \> \> $\boldsymbol b_D\,:=\,$remove $p_{i-1}$ and move $p_i$
 (or $p_i-1$ if $p_i$ is even) to
 the 1st \\
 \> \> \hspace{1cm}  position in $\boldsymbol b_D$; \\
 \> \> $\nu_D\,:=\,$Schreier vector of $K_D$ with respect to $\boldsymbol b_D$;
 \\
 \> \> $\Delta_p\,:=\,p_i^{<K_D>}$; \\
 \\
 \> \>  for each $L$ in ALPHA do\\
 \> \> \> $s\,:=\,$1st element of TAB($L$); \\
 \> \> \> $d\,:=\,$2nd element of TAB($L$); \\
 \> \> \> NEXT$\,:=\,(\Delta_b)^s\cap \Delta_p^{(gd)^{-1}}$; \\
 \\
 \> \> \>  for each $j$ in NEXT do  \\
 \> \> \> \> $s_1\,:=\,trace(j^{s^{-1}},\nu_S)\times s$ ; \\
 \> \> \> \> $d_1\,:=\,$ $d\times trace(j^{g\,d},\nu_D)^{-1}$ ; \\
 \> \> \> \> $L_1\,:=\,$ append $j$ to $L$; \\
 \> \> \> \> TAB($L_1$)\,:=\,$(s_1,d_1)$; \\
 \> \> \> end for; \\
 \> \> \> clear(TAB($L$)); \\
 \> \> end for; \\
 \\
 \> \> ALPHA\,:=\,indices of TAB that were assigned; \\
 \\
 \> \> $(*$ Verify if there are 2 equal permutations of opposite sign in
 $S\times g\times D$ $*)$\\
 \> \> if either $K_S$ or $K_D$ has some permutation with $-1$ then \\
 \> \> \> $(*$ Calculate $sgd$ for all $(s,d)$ in TAB $*)$\\
 \> \> \> set\_sgd$\,:=\,$map $F_2$ on each entry of ALPHA passing TAB \\
 \> \> \> \> \> \hspace{0.7cm} and $g$ as extra arguments; \\
 \> \> \> if set\_sgd has two equal permutations with opposite sign then \\
 \> \> \> \> \> break the loop and \\
 \> \> \> \> \> {\bf return} 0; \\
 \> \> \> end if; \\
 \> \> end if; \\
 \\
 \> \> $(*$ Find the stabilizers $S^{(i+1)}$ and $D^{(i+1)}$ $*)$\\
 \> \> $K_S\,:=\,$remove permutations that have point $b_i$ from $K_S$; \\
 \> \> $K_D\,:=\,$remove permutations that have point $p_i$ from $K_D$; \\
 \\
 \> end for; \\
 \\
 \> $\bar{g}\,:=\,F_2$(any entry of ALPHA); \\
 \> {\bf return} $\bar{g}$; \\
 {\bf end} \\
 \end{tabbing}

 \centerline{\bf Sub routine $F_1$}

 \begin{tabbing}
 \={\bf procedure} \ $F_1$\,($L$,\,TAB,\,$\Delta_D$,\,$\Delta_b$,\,$g$)\\
 \\
 {\bf input:} \= $L$ some entry of ALPHA;\\
 \> TAB table of elements $(s,d)$;\\
 \> $\Delta_D$ all orbits of $K_D$;\\
 \> $\Delta_b$ orbit of some $b_i$; and\\
 \> $g=(\epsilon_\pi,\pi)\in H\otimes S_{2n}$. \\
 \\
 {\bf output:} \= $((\Delta_b)^{sgd})^{<K_D>}$, where $s$ and $d$ are the
 permutations
 associated with $L$. \\
 \> TAB($L$) yields $s$ and $d$.\\
 \hspace{0.5cm}\=\hspace{0.9cm}\=\hspace{0.9cm}\=\hspace{0.9cm}\=\hspace{0.9cm}\=
 \\
 {\bf begin} \\
 \> $sgd\,:=\,F_2\,(L,\,$TAB$,\,g$);\\
 \> {\it result} := \=select the points of all partitions of $\Delta_D$
 that \\
 \> \> have at least one point in $(\Delta_b)^{sgd}$; \\
 \> {\bf return} {\it result}; \\
 {\bf end} \\
 \end{tabbing}

 \centerline{\bf Sub routine $F_2$}

 \begin{tabbing}
 \={\bf procedure} \ $F_2$\,($L$,\,TAB,\,$g$)\\
 \\
 {\bf input:} \= $L$ some entry of ALPHA;\\
 \> TAB table of elements $(s,d)$;\\
 \> $g=(\epsilon_\pi,\pi)\in H\otimes S_{2n}$. \\
 \\
 {\bf output:} \= $sgd$, where $s$ and $d$ are the permutations associated
 with $L$.\\
 \hspace{0.5cm}\=\hspace{0.5cm}\=\hspace{0.5cm}\=\hspace{0.5cm}\=\hspace{0.5cm}\=
 \\
 {\bf begin} \\
 \> $s\,:=\,$ 1st element of TAB($L$); \\
 \> $d\,:=\,$ 2nd element of TAB($L$); \\
 \> {\it result}$\,:=\,s\times g\times d$; \\
 \> {\bf return} {\it result}; \\
 {\bf end} \\
 \end{tabbing}

 \section{Complexity}

 The complexity of the general algorithm to find double
 coset canonical representative is known to be exponential
 in the worst case \cite{Butler2,Hoffmann}.
 On the other hand, the symmetries of tensor expressions are special cases
 of subgroups of $H\otimes S_{2n}$, and actual verifications
 show that in practical applications the algorithm is efficient.
 The symmetries of the Riemann tensor are one of the most complex
 that occur in practice. Therefore, monomials built out
 of Riemann tensors are examples of complex tensor expressions.
 We have implemented algorithm Canonical and the auxiliary routines
 in Maple system \cite{Maple} and have
 developed a program that generates at random Riemann
 monomials of any degree (number of Riemann tensors) with
 all indices contracted (Riemann scalar invariants). For
 each Riemann monomial we calculate the timing to find the
 canonical representative. We use a PC with a processor of 600MHz.
 The vertical axis of the plot of Fig. 1 is the mean of 50
 timings for each monomial. The horizontal axis is the degree.
 We have eliminated all timings of vanishing results.
 \begin{figure}[htbp]
 \hbox to \hsize
 {\hss
 \psfig{file=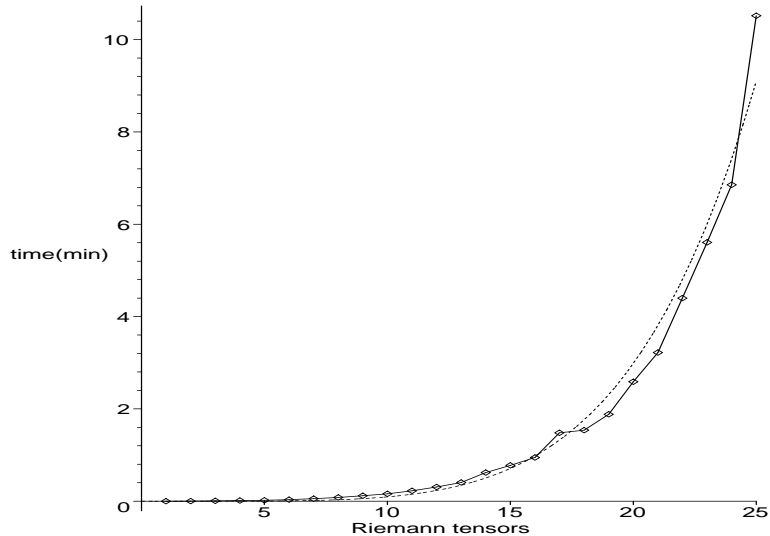,width=10.cm,height=7cm,angle=0}
 \hss}
 \begin{center}
 \begin{minipage}{4.5in}
 \caption{{\footnotesize
 {Timing to find the canonical form of a Riemann monomial
 versus the degree. The dashed line is a fitting
 curve of the form $y=9.3\times 10^{-7}\,x^5$, where $x,y$ are the
 horizontal and vertical axis respectively.}}}
 \end{minipage}
 \end{center}
 \label{Fig.1}
 \end{figure}

 \noindent
 >From Fig. 1 one cannot prove that the algorithm is polynomial.
 It only shows that the implementation in Maple can handle monomials
 with large number of indices. The storage space is
 very low in order to produce the data. If we try to fit the
 experimental curve by a polynomial of the form $y=a\,x^N$,
 for $N<5$ the dashed curve passes above the experimental curve, and
 for $N>5$ the dashed curve passes below for most
 of the points.
 The best polynomial using the least square method
 is $y=9.3\times 10^{-7}\,x^5$.
 Notice that the deviation from the
 polynomial curve depends on the degree due to
 the fact that 50 timings give worse and worse statistics
 with increasing degree.

 \section{Canonicalization of General Expressions}

 Consider an algebraic expression with indexed objects of
 tensorial nature. The product of these objects can be commutative
 or anticommutative. If we expand the expression, it becomes
 a sum of monomials. Refs. \cite{Portugal1} and \cite{Portugal2}
 describe a method for merging monomials into single
 indexed objects, which inherit the symmetries of the original
 objects. The commutative or anticommutative properties are
 converted into permutation symmetries of the merged object.
 At the end, the problem of manipulating an expression
 reduces to the problem of dealing with single
 indexed objects with free and
 dummy indices obeying permutation symmetries.

 Without loss of generality, suppose that the merged object is
 a tensor $T$ with $p$ free indices and $q$ pairs of
 dummy indices. We define the standard configuration as
 \begin{equation}
 T^{i_1 \cdots \, i_{p}} \, ^{d_1}\, _{d_1} \cdots \, ^{d_q}\,_{d_q}.
 \label{abscon}
 \end{equation}
 We do not distinguish contravariant free indices from covariant
 ones. If the original configuration has covariant
 free indices, we pretend that they are contravariant and proceed until
 the end, when the character of the covariant indices is
 restored. This means that there is no preference of putting
 contravariant free indices in front of covariant ones or vice-versa.
 The minimal order is dictated by the base of $S$ which we do not
 know a priori, since it is built out by the
 strong generating set algorithm.
 This choice follows the criteria of least
 computational effort.

 All configurations of (\ref{abscon}) taking into account
 sign changes are given by the application of elements
 in $H\otimes S_{p+2q}$ on (\ref{abscon}).
 Suppose one gives an index configuration.
 The algorithms to canonicalize free and dummy indices
 can be applied in sequence on this configuration.
 The first step is the application of the algorithm of
 ref. \cite{Benar} in order to find the canonical ordering
 and positions of the free indices.
 The next step is the application of the algorithm
 of section 3, translating the points
 $[1,\cdots,2q]$ to the current positions
 of dummy indices. If $[l_1,\cdots,l_{2q}]$ are
 the new positions in increasing order,
 then group $D$ is strongly
 generated by
 \begin{eqnarray}
 \bar{K}_D= \{\,(l_1,l_2),\,\cdots,\,(l_{2q-1},l_{2q}),\hspace{3cm}
 \nonumber\\
 (l_1,l_3)(l_2,l_4),\,\cdots,\,
 (l_{2q-3},l_{2q-1})(l_{2q-2},l_{2q})\}
 \label{KDbar}
 \end{eqnarray}
 with respect to the base $[l_1,l_3,\cdots,l_{2q-1}]$, if the metric is
 symmetric.

 For example, let $R^{abcd}$ be the Riemann tensor and
 we want to canonicalize expression
 \begin{equation}
 R{{_{d_{2}}\,_{d_{3}}\,^{d_{1}}\,^{d_{4}}}}R{{_{d_{5}}\,^b\,^a\,^{d_{2}}}
 }R{{_{d_{4}}\,^{d_{3}}\,_{d_{1}}\,^{d_{5}}}}.
 \label{ex}
 \end{equation}
 Ref. \cite{Portugal1} describes how this
 expression merges into a single tensor, which is
 \begin{equation}
 T{{{_{d_{5}}\,^b\,^a\,^{d_{2}}}
 }\,{_{d_{2}}\,_{d_{3}}\,^{d_{1}}\,^{d_{4}}}}\,
 {{_{d_{4}}\,^{d_{3}}\,_{d_{1}}\,^{d_{5}}}}
 \label{T}
 \end{equation}
 with the following permutations symmetries
 \begin{eqnarray}
 K_S = \{-(1, 2),\, -(3, 4),\, -(5, 6),\, -(7, 8),\,
 -(9, 10),\, -(11, 12),\, \nonumber \\
 (1, 3)(2, 4),\,
 (5, 7)(6, 8),\,
 (9, 11)(10, 12),\, \hspace{2cm} \nonumber \\
 (5, 9)(6, 10)(7, 11)(8, 12)\}. \hspace{3cm}
 \label{KS1}
 \end{eqnarray}
 $K_S$ is a strong generating set.
 The standard configuration is
 \begin{equation}
 T{{{^a\,^b\,^{d_{1}}}}\,{_{d_{1}}\,\cdots\,^{d_{5}}\,_{d_{5}}}}.
 \label{Tstan}
 \end{equation}
 The element of
 $H\otimes S_{12}$, which acts on the standard configuration
 (\ref{Tstan}) and yields (\ref{T}), is
 \begin{equation}
 {g_1} = (1, 12, 11, 4, 5, 6, 8, 9, 10, 7, 3).
 \label{g}
 \end{equation}
 Now we call the algorithm Canonical for free indices (ref. \cite{Benar})
 with the following input: ${g_1}$, $K_S$, and $\boldsymbol b_S=[1,\cdots,11]$.
 We are using the simplest base in order to help the visualization of
 the order of the indices, and we are aware that it has
 unnecessary points.
 The output of the algorithm is
 \begin{equation}
 {g_2}=-(2, 5, 6, 8, 9, 10, 7, 3)(4, 12, 11),
 \label{gbar}
 \end{equation}
 which corresponds to
 \begin{equation}
 -T{^a\,^{d_{2}}\,^b\,{_{d_{5}}}
 }\,{{_{d_{2}}\,_{d_{3}}\,^{d_{1}}\,
 ^{d_{4}}}}\,{{_{d_{4}}\,^{d_{3}}\,_{d_{1}}\,^{d_{5}}}}.
 \label{T1}
 \end{equation}
 The free indices are in the canonical positions, which are
 given by
 \begin{equation}
 [1,2]^{{\,g_2}\,^{-1}}=[1,3],
 \label{f12}
 \end{equation}
 and the positions of dummy indices are
 \begin{equation}
 [3,\cdots,12]^{{\,g_2}\,^{-1}}=[7,11,2,5,10,6,8,9,12,4].
 \label{d12}
 \end{equation}
 Sorting with respect to the basis and concatenating
 (\ref{f12}), (\ref{d12});
 converting to disjoint cycle notation we obtain
 \begin{equation}
 h=-(2, 3),
 \label{h}
 \end{equation}
 which is the group element that converts $K_D$ given by
 (\ref{e4}) to $\bar{K}_D$ given by (\ref{KDbar}) via
 conjugation, i.e. $\bar{K}_D=h^{-1}\times K_D\times h$.
 The input of the algorithm Canonical for dummy indices
 (section 3) is
 \begin{eqnarray}
 K_{S_{1,3}} = \{-(5, 6),\, -(7, 8),\,
 -(9, 10),\, -(11, 12),\,  \hspace{2cm} \nonumber \\
 (5, 7)(6, 8),\, (9, 11)(10, 12),\,
 (5, 9)(6, 10)(7, 11)(8, 12)\},
 \label{KS1bar}
 \end{eqnarray}
 \begin{equation}
 \bar{\boldsymbol b}_S=[2,4,5,\cdots,11],
 \end{equation}
 and
 \begin{equation}
 {g_3}={g_2}\,h=-(2, 5, 6, 8, 9, 10, 7)(4, 12, 11).
 \label{gbarbar}
 \end{equation}
 The algorithm must be modified so that
 the generating set for group $D$ must be
 \begin{eqnarray}
 \bar{K}_D = \{(2, 4),\,
 (5, 6),\,
 (7, 8),\,
 (9, 10),\,
 (11, 12),\hspace{2cm}\nonumber \\
 (2, 5)(4, 6),\,
 (5, 7)(6, 8),\,
 (7, 9)(8, 10),\,
 (9, 11)(10, 12)\},
 \label{KSD1bar}
 \end{eqnarray}
 with base $\bar{\boldsymbol b}_D=[2,5,7,9,11]$.
 The output is
 \begin{equation}
 {g_4}=-(4, 5)(6, 7, 9)(8, 11).
 \label{gbarbarbar}
 \end{equation}
 The permutations ${{g_3}}$ and
 ${g_4}$ do not act on the
 standard configuration (\ref{Tstan}). They act on
 \begin{equation}
 T^a\,^{d_1}\,^b\,_{d_1}\,^{d_2}\,_{d_2}\, \cdots \,^{d_5}\,_{d_5}.
 \label{abscon1}
 \end{equation}
 The final answer is
 \begin{equation}
 g_5={g_4}\,h^{-1}=-(2, 3)(4, 5)(6, 7, 9)(8, 11).
 \label{can_g}
 \end{equation}
 In terms of tensor notation, the canonical form is
 \begin{equation}
 -R{^a\,^{d_{1}}\,^b\,{^{d_{2}}}R{{_{d_{1}}\,^{d_{3}}\,^{d_{4}}\,^{d_{5}}}}
 }R{{_{d_{2}}\,_{d_{4}}\,_{d_{3}}\,_{d_{5}}}},
 \label{final_ex}
 \end{equation}
 which is obtained acting $g_5$ on (\ref{Tstan}) and splitting
 back the merged tensor.

 \

 \noindent {\bf Acknowledgments}

 \

 \noindent
 We thank Drs. A. Hulpke and G. A\~na\~nos for discussions on this subject.

 \

 \end{document}